\documentclass[12pt,preprint]{aastex}

\shorttitle{$\kappa$ CrB NPOI Observations}
\shortauthors{Baines et al.}

\begin{document}

\title{NPOI Observations of the Exoplanet Host $\kappa$ Coronae Borealis and Their Implications for the Star's and Planet's Masses and Ages}

\author{Ellyn K. Baines, J. Thomas Armstrong}
\affil{Remote Sensing Division, Naval Research Laboratory, 4555 Overlook Avenue SW, \\ Washington, DC 20375}
\email{ellyn.baines@nrl.navy.mil}

\author{Gerard T. van Belle}
\affil{Lowell Observatory, Flagstaff, AZ 86001} 

\begin{abstract}

We used the Navy Precision Optical Interferometer to measure the limb-darkened angular diameter of the exoplanet host star $\kappa$ CrB and obtained a value of 1.543$\pm$0.009 mas. We calculated its physical radius (5.06$\pm$0.04 $R_\odot$) and used photometric measurements from the literature with our diameter to determine $\kappa$ CrB's effective temperature (4788$\pm$17 K) and luminosity (12.13$\pm$0.09 $L_\odot$). We then placed the star on an H-R diagram to ascertain the star's age (3.42$^{\rm +0.32}_{\rm -0.25}$ Gyr) and mass (1.47$\pm$0.04 $M_\odot$) using a metallicity of [Fe/H] = +0.15. With this mass, we calculated the system's mass function with the orbital elements from a variety of sources, which produced a range of planetary masses: $m_{\rm p} \sin i$ = 1.61--1.88 $M_{\rm Jup}$. We also updated the extent of the habitable zone for the system using our new temperature.

\end{abstract}



\section{Introduction} 
$\kappa$ Coronae Borealis ($\kappa$ CrB, HD 142091, HR 5901, FK5 1414, HIP 77655) is a ``retired A star'', otherwise known as a former A-type dwarf that is now a K-type subgiant. A planetary companion was discovered by \citet{2008ApJ...675..784J} with a minimum mass $m_{\rm p} \sin i$ of 1.8 $M_{\rm Jup}$, a separation $a$ of 2.7 AU, an orbital eccentricity $e$ of 0.146$\pm$0.08, and a period $P$ of 1208$\pm$30 days. Johnson et al. used \citet{2002AandA...391..195G} theoretical mass tracks to determine the star's mass and age: $M_\star = 1.80\pm0.11$ $M_\odot$ and age = 2.5$\pm$1.0 Gyr. Johnson et al. also used spectral analysis to determine that $\kappa$ CrB is metal rich ([Fe/H]=+0.15$\pm$0.05), has an effective temperature $T_{\rm eff}$ of 4970$\pm$70 K, and has a surface gravity log $g$ of 3.47$\pm$0.09 cm s$^{\rm -2}$. Finally, they estimated the luminosity $L$ and radius $R$ to be $12.3 \pm 0.04$ $L_\odot$ and $4.71 \pm 0.08$ $R_\odot$, respectively.


More recently, \citet{2013arXiv1302.7000B} showed spatially resolved images of the dust belt or belts around $\kappa$ CrB with two possible configurations: a single wide belt spanning 20 to 220 AU or two narrow belts at $\sim$40 AU and 165 AU. They also announced the presence of a second companion, though the nature of that companion is not yet determined. If the companion resides between the wide belt and the host star, it has a mass $>$ 5 $M_{\rm Jup}$ and is therefore an exoplanet. If the companion resides between the two narrow belts, it has a larger mass of $>$ 13 $M_{\rm Jup}$ and is a brown dwarf instead.


In order to more fully characterize the star and its known exoplanet, we measured the angular diameter of $\kappa$ CrB using the Navy Precision Optical Interferometer (NPOI). We calculated its radius and $T_{\rm eff}$ and used these values with a stellar model to place the star on an Hertzsprung-Russell (H-R) diagram so we could determine its mass and age. We calculated the mass function using the orbital elements from a number of sources to ascertain the planet's mass. Section 2 discusses our observing process; Section 3 describes the visibility measurements and how we calculated stellar parameters such as angular diameter, effective temperature, physical radius, and luminosity; Section 4 explores the extent of $\kappa$ CrB's habitable zone as well as the mass and age of the star and its planet; and Section 5 summarizes our findings.


\section{Interferometric Observations}

We observed $\kappa$ CrB using the NPOI, an interferometer located on Anderson Mesa, AZ \citep{1998ApJ...496..550A}. The NPOI consists of two nested arrays: the four stations of the astrometric array (astrometric center, east, west, and north) and the six stations of the imaging array, of which two stations (E6 and W7) are currently in operation and three more will be coming online in the near future (E7, E10, and W10). The current baselines, i.e., the distance between the stations, range from 16 to 79 m, though our maximum baseline will be 432 m when the E10 and W10 stations are completed within the next year. 
We use the central 12 cm of the 50-cm siderostats and observe in 16 spectral channels spanning 550 to 850 nm simultaneously.

Our observing procedure and data reduction process are described in detail in \citet{2003AJ....125.2630H}. To summarize, each observation consisted of a scan on the fringe (a coherent scan) where the fringe contrast was measured every 2 ms paired with an off the fringe scan (an incoherent scan) subsequently used to estimate the biases affecting our measurements. We obtained scans on three baselines at a time. Each coherent scan was averaged to 1-second data points and then to a 30-second average. The internal errors were estimated using the dispersion of the 1-second data points.

We observed $\kappa$ CrB for three nights in March 2013 using baseline lengths from 19 to 53 m. We interleaved data scans of $\kappa$ CrB with two calibrators -- HD 143894 and HD 147394 -- and then converted the instrumental target and calibrator visibilities to calibrated visibilities for the target. We chose HD 143894 and HD 147394 as calibrator stars because they are nearly unresolved on the baselines used and therefore mimic point sources. This minimizes error introduced into the calibrated data that arise from uncertainties in the calibrator stars' angular diameters, and the errors in those diameter estimates are taken into account in the calibration process.

We estimated the calibrator stars' sizes by constructing their spectral energy distribution (SED) fits using photometric values published in \citet{1965ArA.....3..439L}, \citet{1981PDAO...15..439M}, \citet{1993AandAS..102...89O}, \citet{1990VilOB..85...50J}, \citet{1972VA.....14...13G}, \citet{1970AandAS....1..199H}, \citet{1991TrSht..63....1K}, \citet{1968tcpn.book.....E}, \citet{1966CoLPL...4...99J}, \citet{2003tmc..book.....C}, and \citet{1993cio..book.....G} as well as spectrophotometry from \citet{1983TrSht..53...50G}, \citet{1998yCat.3207....0G}, \citet{1997yCat.3202....0K} obtained via the interface created by \citet{1997AandAS..124..349M}. The assigned uncertainties for the 2MASS infrared measurements are as reported in \citet{2003tmc..book.....C}, and an error of 0.05 mag was assigned to the optical measurements. We determined the best fit stellar spectral template to the photometry from the flux-calibrated stellar spectral atlas of \citet{1998PASP..110..863P} using the $\chi^2$ minimization technique. The resulting diameter estimates are 0.39$\pm$0.02 mas for HD 143894 and 0.36$\pm$0.08 mas for HD 147394. 

Another parameter that can affect interferometric measurements is calibrator star oblateness. In this case, the effects are negligible. HD 143894 has a $v \sin i$ of 115 to 138 km s$^{\rm -1}$ \citep{1995ApJS...99..135A,2003AandA...398.1121E} that may produce some oblateness on the order of 3--4$\%$. However, because the star is small, the effect of even a 4$\%$ variation across the projected disk radius does not significantly affect the data calibration. It changes the measured diameter by less than 1$\%$, and even the 4$\%$ variation is within the error bars of the calibrator size. HD 147394 with a $v \sin i$ of 46 km s$^{\rm -1}$ at most shows even less oblateness at 1$\%$ \citep{2002AandA...393..897R}.


\section{Results}

\subsection{Angular Diameter Measurement}

Interferometric diameter measurements use $V^2$, the square of the fringe visibility. For a point source, $V^2$ is unity, while for a uniformly-illuminated disk, $V^2 = [2 J_1(x) / x]^2$, where $J_1$ is the Bessel function of the first order, $x = \pi B \theta_{\rm UD} \lambda^{-1}$, $B$ is the projected baseline toward the star's position, $\theta_{\rm UD}$ is the apparent UD angular diameter of the star, and $\lambda$ is the effective wavelength of the observation \citep{1992ARAandA..30..457S}. We obtained $\theta_{\rm UD} = 1.449\pm0.009$ mas.\footnote{The imaged dust disk does not affect our observed visibilities. Even at the smallest possible separation, the disk's angular distance from $\kappa$ CrB means it is outside the NPOI's field of view. Additionally, the brightness difference between the star and disk is beyond the NPOI's dynamical range.} Our data files in OIFITS format are available upon request.

\begin{figure}[h]
\includegraphics[width=0.75\textwidth, angle=90]{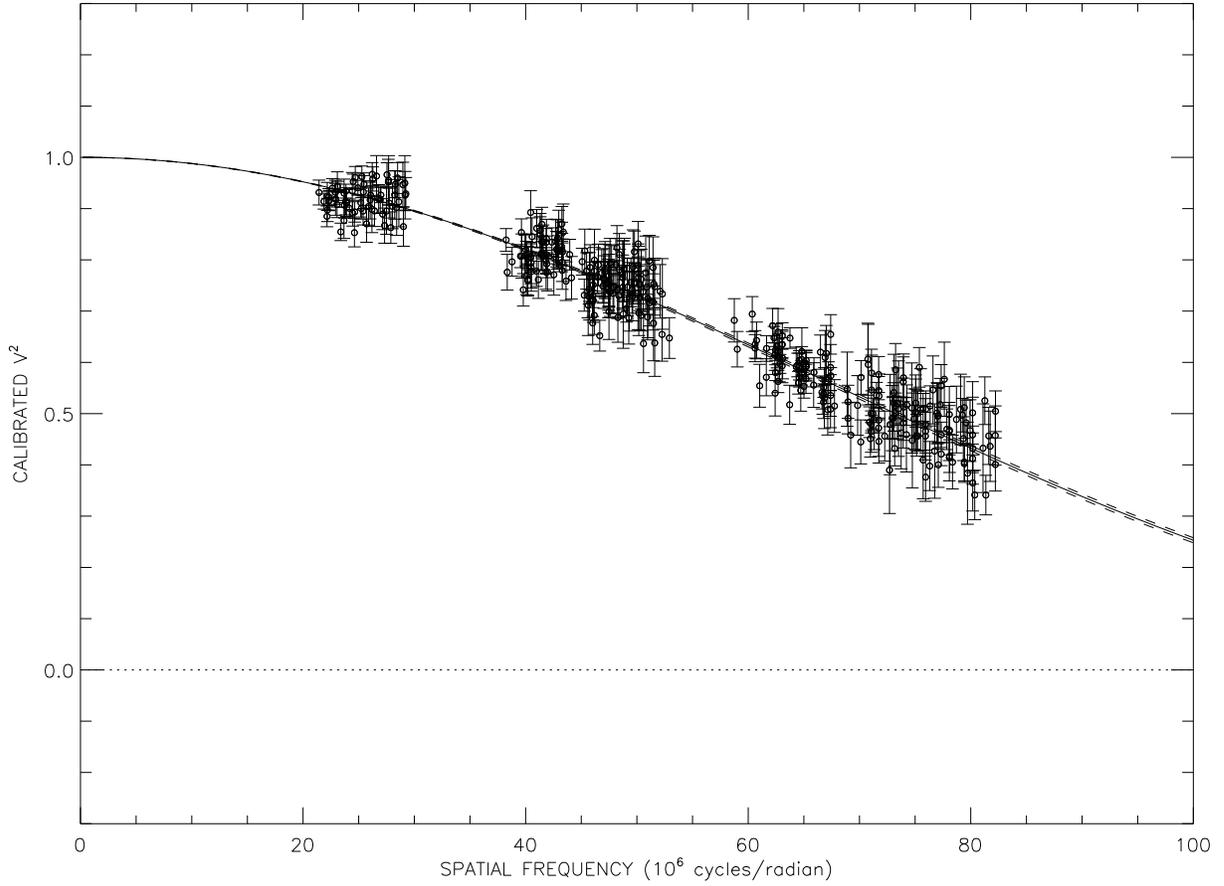}
\caption{$\kappa$ CrB $\theta_{\rm LD}$ fit. The solid line represents the theoretical visibility curve for the best fit $\theta_{\rm LD}$, the dashed lines are the 1$\sigma$ error limits of the diameter fit, the open circles are the calibrated visibilities, and the vertical lines are the measurement errors.}
  \label{ldplot}
\end{figure}

A more realistic model of a star's disk includes limb darkening (LD).  If a linear LD coefficient $\mu_\lambda$ is used,
\begin{equation}
V^2 = \left( {1-\mu_\lambda \over 2} + {\mu_\lambda \over 3} \right)^{-1}
\times
\left[(1-\mu_\lambda) {J_1(\rm x_{\rm LD}) \over \rm x_{\rm LD}} + \mu_\lambda {\left( \frac{\pi}{2} \right)^{1/2} \frac{J_{3/2}(\rm x_{\rm LD})}{\rm x_{\rm LD}^{3/2}}} \right] .
\end{equation}
where x$_{\rm LD} = \pi B\theta_{\rm LD}\lambda^{-1}$ \citep{1974MNRAS.167..475H}.  To estimate $\mu_\lambda$, we made initial estimates, $T_{\rm eff}$ = 4796 K and log $g$ = 3.29 cm s$^{\rm -2}$, from averaging over the 20 citations listed on VizieR, and then used \citet{2011AandA...529A..75C} to obtain $\mu_\lambda$ = 0.68. Our resulting value for $\theta_{\rm LD}$ is 1.543$\pm$0.009 mas. Figure \ref{ldplot} shows the $\theta_{\rm LD}$ fit to $V^2$ for $\kappa$ CrB. 

The error for the LD diameter fit was derived using the method described in \citet{2010SPIE.7734E.103T}. Tycner et al. demonstrated that a non-linear least-squares method does not adequately take into account the atmospheric effects on time scales briefer than the minutes between the target and calibrator observations. They instead propose a bootstrap Monte Carlo method that retains the inherent structure of the observations, i.e., groups of data points, instead treating them as individual data points. This is because the NPOI scans across 16 channels simultaneously and collects data in groups. Tycner et al. found that when the data points were analyzed individually, the deviation of a single scan from the overall trend had a significant impact on the resulting diameter and error calculation. But when they used scans of 16 channels as a group instead of as single data points, the uncertainty on the angular diameter determination was larger and more realistic than methods that rely solely on formal errors. 

\subsection{Stellar Radius, Luminosity and Effective Temperature}

$\kappa$ CrB has a parallax of 32.79$\pm$0.21 mas \citep{2007hnrr.book.....V}, which translates to a distance of 30.50$\pm$0.20 pc. We combined this with our measured $\theta_{\rm LD}$ to calculate the physical radius of the star: 5.06$\pm$0.04 $R_\odot$. 

In order to determine the $L$ and $T_{\rm eff}$ of $\kappa$ CrB, we constructed its SED using the sources and technique of fitting spectral templates to observed photometry as described in Section 2. The resulting SED gave us the bolometric flux ($F_{\rm BOL}$) and allowed for extinction $A_{\rm V}$ with the wavelength-dependent reddening relations of \citet{1989ApJ...345..245C}. The K0.5 III template provided the best fit with a $F_{\rm BOL}$ of 4.17$\pm$0.04 $\times 10^{\rm -7}$ erg s$^{\rm -1}$ cm$^{\rm -2}$ and $A_{\rm V}$ of 0.03$\pm$0.01 magnitudes.

We combined our $F_{\rm BOL}$ with $\kappa$ CrB's distance to estimate its luminosity using $L = 4 \pi d^2 F_{\rm BOL}$, which produced a value of 12.13$\pm$0.19 $L_\odot$. We also combined the $F_{\rm BOL}$ with $\theta_{\rm LD}$ to determine $\kappa$ CrB's effective temperature by inverting the relation,
\begin{equation}
F_{\rm BOL} = {1 \over 4} \theta_{\rm LD}^2 \sigma T_{\rm eff}^4,
\end{equation}
where $\sigma$ is the Stefan-Bolzmann constant ($5.6704 \times 10^{\rm -5}$ erg cm$^{\rm -2}$ s$^{\rm -1}$ deg$^{\rm -4}$) and $\theta_{\rm LD}$ is in radians. This produces an effective temperature of $4788 \pm 17$ K. Because $\mu_\lambda$ is chosen based on a given $T_{\rm eff}$, we checked to see if $\mu_\lambda$ would change based on our new $T_{\rm eff}$. $\mu_\lambda$ remained the same, as the temperature difference between the literature's average and our value was only 8 K. All of these results are listed in Table \ref{results}.


\section{Discussion}

\subsection{Habitable Zone of $\kappa$ CrB}

$\kappa$ CrB is currently the only subgiant star with resolved images of a dust disk and is an unusual case because it is an intermediate mass star boasting both planets and planetesimal belts \citep{2013arXiv1302.7000B}. We wanted to determine the habitable zone (HZ) of $\kappa$ CrB to see if it overlapped either of the dust disks or either companion. We used the following equations from \citet{2006ApJ...649.1010J}:
\begin{equation}
S_{b,i}(T_{\rm eff}) = (4.190 \times 10^{-8} \; T_{\rm eff}^2) - (2.139 \times 10^{-4} \; T_{\rm eff}) + 1.296
\end{equation}
and 
\begin{equation}
S_{b,o}(T_{\rm eff}) = (6.190 \times 10^{-9} \; T_{\rm eff}^2) - (1.319 \times 10^{-5} \; T_{\rm eff}) + 0.2341
\end{equation}
where $S_{b,i}$($T_{\rm eff}$) and $S_{b,o}$($T_{\rm eff}$) are the critical fluxes at the inner and outer boundaries of the HZ in units of the solar constant. The inner and outer physical boundaries $r_{i,o}$ in AU were then calculated using
\begin{equation}
r_i = \sqrt{ \frac{L/L_\odot}{S_{b,i}(T_{\rm eff})} } \; \; \; \; \; {\rm and} \; \; \; \; \; r_o = \sqrt{ \frac{L/L_\odot}{S_{b,o}(T_{\rm eff})} }.
\end{equation}
We obtained habitable zone boundaries of 3.05 AU and 6.06 AU. $\kappa$ CrB's inner planet has semimajor axis of 2.80 AU \citep{2010ApJ...709..396B}, just inside the inner boundary. The outer companion is beyond the outer limit even if the configuration with the smallest separation of 7.2 AU from Bonsor et al. holds true. 

\subsection{Mass and Age of $\kappa$ CrB and its Planet}

\begin{figure}[h]
\includegraphics[width=0.75\textwidth, angle=90]{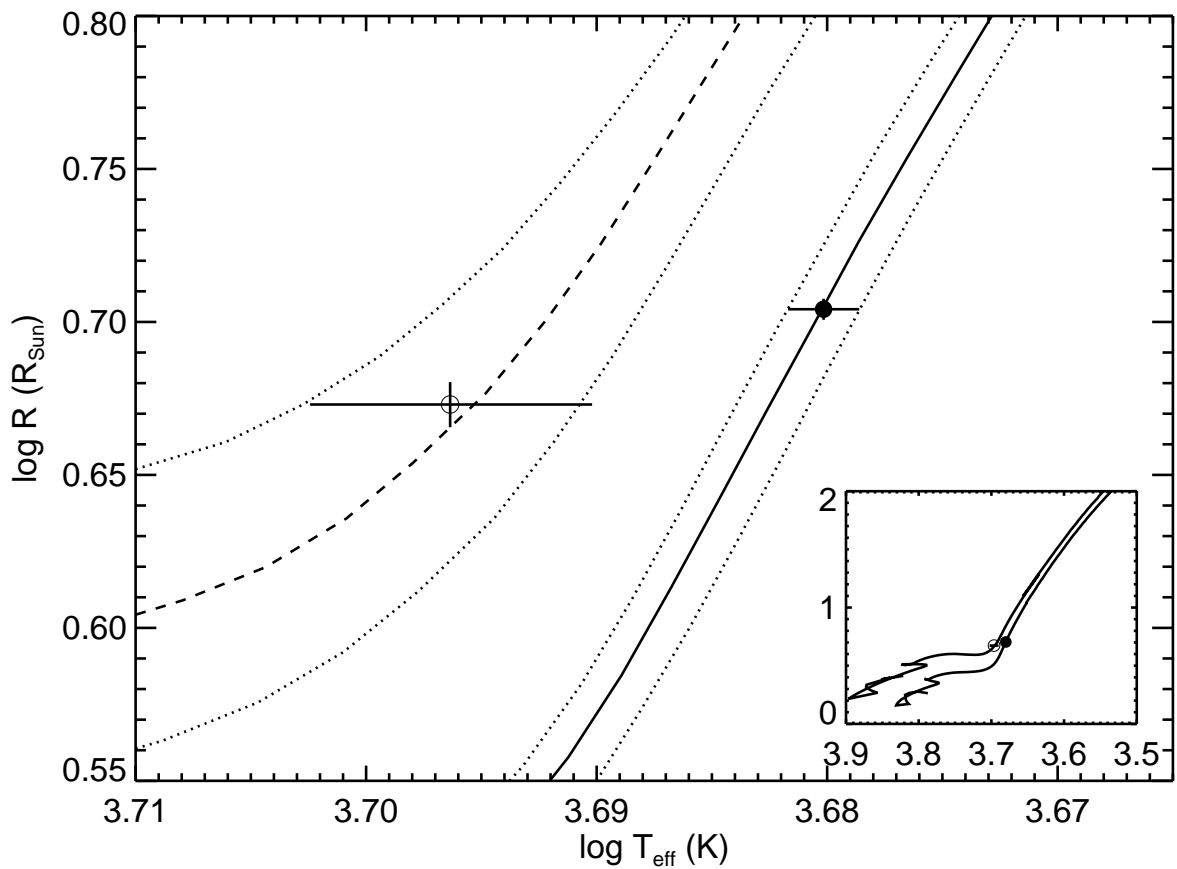}
\caption{Radius and temperature of $\kappa$ CrB plotted along with Y$^2$ mass tracks \citep{2001ApJS..136..417Y} using [Fe/H]=+0.15 \citep{2008ApJ...675..784J}. The open circle is Johnson et al.'s data point, and the dashed line is the 1.80 $M_\odot$ track. The filled circle represents our $R$ and $T_{\rm eff}$ and the solid line corresponds to a mass track of 1.47 $M_\odot$. The dotted lines are the errors in the masses, $\pm$0.11 $M_\odot$ for Johnson et al. and $\pm$0.04 $M_\odot$ for our data. The inset plot shows the full evolution track.}
  \label{hr}
\end{figure}

We used our newly calculated $T_{\rm eff}$ and $R$ to place the star on an H-R diagram using stellar evolutionary models from \citet{2001ApJS..136..417Y} to determine its mass and age. Our best fit mass is 1.47$\pm$0.04 $M_\odot$, which is an uncertainty of 3$\%$ (see Figure \ref{hr}). This corresponds to an age of 3.42$^{\rm +0.32}_{\rm -0.25}$ Gyr, which is older than the age determined by \citet{2008ApJ...675..784J} of 2.5$\pm$1.0 Gyr, though technically still within the errors. Our stellar mass is lower than Johnson et al.'s, who listed a mass of 1.80$\pm$0.11 $M_\odot$, though it agrees with $M_\star$ determined by other groups, namely 1.51 $M_\odot$ \citep{2008PASJ...60..781T,2012PASJ...64..135S}.

The literature also includes a range of metallicities [Fe/H] that span --0.10 \citep{2007ApJS..171..146S} to +0.20 \citep{2010AandA...515A.111S}. This range had a substantial effect on the resulting stellar masses: $M_\star$ = 1.15 $M_\odot$ for the lowest [Fe/H] and $M_\star$ = 1.51 $M_\odot$ for the highest. For our final $M_\star$, we used Johnson et al.'s [Fe/H] of +0.15 because their work was our primary source of comparison for $\kappa$ CrB's mass and age. 

We used our best fit mass and the orbital elements from the four different sources listed in Table \ref{orbit} to calculate the mass function 
\begin{equation}
f(m) = \frac{(m_{\rm p} \sin i)^3}{(M_{\star} + m_{\rm p})^2} = \frac{P}{2 \pi G} (K \sqrt{1-e^2})^3,
\end{equation}
where $G$ is the gravitational constant and then calculated the planet's mass. This produced a range of $m_{\rm p} \sin i$ from 1.61$\pm$0.10 to 1.88$\pm$0.09 $M_{\rm Jup}$, depending on which orbital elements were used, and all masses converged in two iterations. The values are listed in Table \ref{orbit}.


\section{Summary}

$\kappa$ CrB remains a fascinating target because of its unusual configuration. It has at least one exoplanet and perhaps two, and features a single wide dust ring or two narrow ones. We observed the star interferometrically in order to help characterize the main star and the environment in which the planet(s) and dust ring(s) reside.

We determined a variety of fundamental parameters for the retired A star $\kappa$ CrB: the limb-darkened angular diameter (1.543$\pm$0.009 mas), the physical size (5.06$\pm$0.04 $R_\odot$), the effective temperature (4788$\pm$17 K), the luminosity (12.13$\pm$0.19 $L_\odot$), the mass (1.47$\pm$0.04 $M_\odot$), age (3.42$^{\rm +0.32}_{\rm -0.25}$ Gyr), and habitable zone range (3.05 to 6.06 AU). We combined our mass with the orbital parameters from four sources to calculate the planet's mass and have a range of $m_{\rm p} \sin i$ = 1.61 to 1.88 $M_{\rm Jup}$.

\acknowledgments

The Navy Precision Optical Interferometer is a joint project of the Naval Research Laboratory and the U.S. Naval Observatory, in cooperation with Lowell Observatory, and is funded by the Office of Naval Research and the Oceanographer of the Navy. This research has made use of the SIMBAD database, operated at CDS, Strasbourg, France. This publication makes use of data products from the Two Micron All Sky Survey, which is a joint project of the University of Massachusetts and the Infrared Processing and Analysis Center/California Institute of Technology, funded by the National Aeronautics and Space Administration and the National Science Foundation.

\clearpage

\begin{deluxetable}{lcl}
\tablewidth{0pc}
\tablecaption{$\kappa$ CrB Stellar Parameters.\label{results}}
\tablehead{ \colhead{Parameter} & \colhead{Value} & \colhead{Reference} }
\startdata
\cline{1-3}
\cline{1-3}
\multicolumn{3}{c}{From the literature:} \\
$\pi$ (mas) & 32.79$\; \pm \;$0.21 & \citet{2007hnrr.book.....V} \\
Distance (pc) & 30.50$\; \pm \;$0.20 & Calculated from $\pi$ \\
$\mu_{\lambda}$ & 0.68 & \citet{2011AandA...529A..75C} \\
\cline{1-3}
\cline{1-3}
\multicolumn{3}{c}{The results of our SED fit:} \\
$A_{\rm V}$ (mag)   & 0.03$\; \pm \;$0.01 & \\
$F_{\rm BOL}$ (10$^{-7}$ erg s$^{-1}$ cm$^{-2}$) & 4.17$\; \pm \;$0.04 & \\
\cline{1-3}
\cline{1-3}
\multicolumn{3}{c}{The results of this work:} \\
$\theta_{\rm UD}$ (mas) & 1.449$\; \pm \;$0.009 & \\
$\theta_{\rm LD}$ (mas) & 1.543$\; \pm \;$0.009 &  \\
$\theta_{\rm LD}$ fit reduced $\chi^2$ (392 points) & 1.2 &  \\
$R_{\rm linear}$ ($R_\odot$) &  5.06$\; \pm \;$0.04 &  \\
$T_{\rm eff}$ (K) & 4788$\; \pm \;$17 &  \\
$L$ ($L_\odot$) & 12.13$\; \pm \;$0.19  &  \\
Mass ($M_\odot$) & 1.47 $\pm$ 0.04 &  \\
Age (Myr) & 3.42 $^{\rm +0.32}_{\rm -0.25}$ &  \\
\enddata
\end{deluxetable}

\clearpage

\begin{deluxetable}{lcccc}
\tablewidth{0pc}
\tablecaption{$\kappa$ CrB Orbital Elements and $M_{\rm planet}$.\label{orbit}}
\tablehead{\colhead{Parameter} & \colhead{J08} & \colhead{B10} &\colhead{S12} & \colhead{B13} }
\startdata
$P$ (days)          & 1208$\pm$30 	& 1261.94$^{+28.91}_{-23.97}$ & 1251$\pm$15     & 1300$\pm$15 \\
$e$                 & 0.146$\pm$0.08 	& 0.044 ($<$0.123)            & 0.073$\pm$0.049 & 0.125$\pm$0.049 \\
$K$ (m s$^{\rm -1}$)  & 24.0$\pm$1        & 25.17$^{+1.12}_{-1.55}$     & 23.6$\pm$1.1    & 27.3$\pm$1.3 \\
$a$ (AU)            & 2.7 		& 2.80$^{+0.07}_{-0.08}$      & 2.6             & 2.8$\pm$0.1 \\
$m_{\rm P} \sin i$ ($M_{\rm Jup}$) & 1.8 & 2.01$^{+0.11}_{-0.17}$     & 1.6             & 2.1 \\
\cline{1-5}
\cline{1-5}
\multicolumn{5}{c}{The results of this work combined with above elements:} \\
$m_{\rm p} \sin i$ ($M_{\rm Jup}$) & 1.61$\pm$0.10 & 1.73$\pm$0.04 & 1.61$\pm$0.12 & 1.88$\pm$0.09 \\
$m_{\rm p}$ ($M_{\rm Jup}$)$^\dagger$ & 1.86 & 2.00 & 1.86 & 2.17 \\
\enddata
\tablecomments{$^\dagger$This is the planet's mass assuming the dust disk and planet share the same inclination angle $i_{\rm disk}$=60$^\circ$ (Bonsor et al. 2013);
J08 is Johnson et al. (2008); B10 is Bowler et al. (2010); S12 is Sato et al. (2012); B13 is Bonsor et al. (2013).}
\end{deluxetable}

\clearpage


\clearpage

\end{document}